\headline={\ifnum\pageno>1 \hss \number\pageno\ \hss \else\hfill \fi}
\pageno=1
\nopagenumbers
\hbadness=100000
\vbadness=100000
%----------------------------------------------------------------------------

\centerline{\bf PERMUTATION WEIGHTS FOR AFFINE LIE ALGEBRAS }
\vskip 15mm
\centerline{\bf H. R. Karadayi \footnote{*}{
e-mail: karadayi@itu.edu.tr} and M. Gungormez}
\centerline{Dept.Physics, Fac. Science, Tech.Univ.Istanbul }
\centerline{ 80626, Maslak, Istanbul, Turkey }
\vskip 10mm
\centerline{\bf{Abstract}}
\vskip 10mm

We show that permutation weights, which are previously introduced for finite
Lie algebras, can be appropriately defined also for affine Lie algebras. This
allows us to classify all the weights of an affine Weyl orbit explicitely. Let
$\Lambda$ be a dominant weight of an affine Lie algebra $ G_N^{(r)} $
for r=1,2,3. At each and every order M of weight depths, the set
$\wp_M(\Lambda)$ of permutation weights is formed out of a finite number of
dominant weights of the finite Lie algebra $G_N$. In case of $A_N^{(1)}$
algebras, we give the rules to determine the elements of a $\wp_M(\Lambda)$
completely.

As being a positive test of our proposal, we consider the problem of
calculating weight multiplicities for affine Lie algebras and hence
our discussions are based on explicit computations of Weyl-Kac
character formula. It is known that weight multiplicities are provided
by string functions which are defined to be formal power series
$\sum_{M=0}^\infty C(M) \ q^M$ where the order M specifies the {\bf depth} of
weights contributing to C(M). In the conventional calculational schemes which
are based on Kac-Peterson form of affine Weyl groups, Weyl-Kac formula includes
a sum over a part of the whole root lattice and hence it is seen that the roots
of the same length contribute in general to C(M) for several values of M.
On the contrary, we will determine, for any fixed value of M, the complete set
of weights having depth M and contributing only to C(M).

For applications of Weyl-Kac formula, one must also know the signatures which
correspond to weights within the Weyl orbits of strictly dominant weights.
This is given by the aid of a properly defined index.

Another emphasis is that the way of discussion adopted here gives us a
possibility for extensions to other infinite dimensional Lie algebras beyond
affine Lie algebras.

\vskip 25mm
\vskip 25mm

\hfill\eject

\vskip 3mm
\noindent {\bf{I.\ INTRODUCTION}}
\vskip 3mm

It is well-known that the Weyl group of a Lie algebra plays an essential role
in understanding of the structure of this algebra and also all of its
representations {\bf [1]}. In high energy physics, Weyl groups also enter in
many areas such as calculation of fusion coefficients {\bf [2]} or
construction of exactly solvable models {\bf [3]}. The structure of Weyl
groups is completely known for finite Lie algebras in principle. In practice
however, the problem is not so trivial even for finite Lie algebras, especially
for the ones with some higher rank. The order of $E_8$ Weyl group is, for
instance, 696729600 and any application of Weyl character formula needs for
$E_8$ an explicit calculation of a sum over 696729600 Weyl reflections. We
have shown in a recent work {\bf [4]} that there is a way to overcome this
difficulty.

For infinite dimensional Lie algebras, on the other hand, the situation is
more complex because their Weyl groups are also infinite dimensional. Thanks
to the seminal work of Kac and Peterson {\bf [5]}, it is known that affine Weyl
groups can be expressed in the form of semi-direct products of a kind of
translations and finite Weyl groups. It is commonly known, however, that
there is a lack of knowledge concerning Weyl groups of infinite dimensional
Lie algebras beyond affine Lie algebras.

An efficient way to describe the structure of a Weyl group is to study its
action on the whole weight lattice. In many respects, this requires an
explicit description of weights participating in the Weyl orbits. To this end,
the Weyl character formula {\bf [6]} establishes a principal example.
We will concern ourselves here mainly with the calculation of characters
by the aid of this formula. For a dominant
weight $\Lambda^+$, it is known that the character $Ch(\Lambda^+)$ of its
irreducible representation $R(\Lambda^+)$ is defined by
$$ Ch(\Lambda^+) \equiv \sum_{\lambda^+} \ \sum_{\mu \in W(\lambda^+)}
m_{\Lambda^+}(\mu) \ e^\mu \eqno(I.1) $$
where the first sum is over $\Lambda^+$ and all of its sub-dominant weights
and $m_{\Lambda^+}(\mu)$\ 's are multiplicities which count the number of times
a weight $\mu$ is repeated for $R(\Lambda^+)$. Note here that multiplicities
are invariant under Weyl group actions and hence it is sufficient to determine
only $m_{\Lambda^+}(\lambda^+)$ for the whole Weyl orbit $W(\lambda^+)$.
The crucial point here is however to calculate the multiplicities as positive
integers. For this, the significant formula of Weyl says that
$$ Ch(\Lambda^+) = { A(\Lambda^{++}) \over A(\rho) }  \eqno(I.2) $$
where $\Lambda^{++}$ is the strictly dominant weight defined by
$$ \Lambda^{++} \equiv \rho+\Lambda^+   \eqno(I.3)   $$
and $\rho$ here is the Weyl vector which is also a strictly dominant weight.
The object $A(\Lambda^{++})$ covers by definition a sum over the whole Weyl
group of underlying Lie algebra. As we have shown for finite Lie
algebras {\bf [7]}, such a definition can always be replaced by the
formula
$$ A(\Lambda^{++}) \equiv \sum_{\mu \in W(\Lambda^{++})} \epsilon(\mu) \
e^\mu \eqno(I.4) $$
where $W(\Lambda^{++})$ is the Weyl orbit of $\Lambda^{++}$. Note here that,
in the classical formulation, signatures are defined for Weyl reflections
though, in (I.4),  $\epsilon(\mu)$ is the extended signature which can be
defined properly for any weight $\mu \in W(\Lambda^{++})$. Formal exponentials
$e^{\mu}$ in (I.4) are defined just as in the book of Kac {\bf [8]}.
We will show in this paper that (I.4) is also valid for affine Lie algebras.

For finite Lie algebras, the equality of (I.1) and (I.2) is successful to
bring us the multiplicities as positive integers. It is immediately seen
however that, without a modification, this is not possible for affine
Lie algebras. Let us define {\bf string functions} by
$$ {\it C}_{\Lambda^+}(\mu) \equiv \sum_{M=M_0(\mu,\Lambda^+)}
e^{-\delta M} \ c_{\mu,\Lambda^+}(M)  \eqno(I.5)  $$
where $M_0(\mu,\Lambda^+)$\ is the lowest depth for a sub-dominant weight
$\lambda^+$ for which $ \mu \in W(\lambda^+)$. From discussions above, we
know that $M_0(\Lambda^+,\Lambda^+) \equiv 0$. The basic observation for
affine Lie algebras is to replace, in (I.1), ${\it C}_{\Lambda^+}(\mu)$
instead of $m_{\Lambda^+}(\mu)$. This allows us to calculate the coefficients
$c_{\mu,\Lambda^+}(M)$ as positive integers. With this modification,
it is known that Weyl formula turns to Weyl-Kac formula.

\hfill\eject

As a result of outstanding efforts of Kac and some other authors, it is shown
that such a modification in Weyl character formula provides us weight
multiplicities for a general class of infinite dimensional Lie algebras.
It is also noteworthy that we have another application of Weyl character
formula concerning Borcherd's algebras {\bf [9]}. For all these, it is clear
that we have to know, in the framework of this work, the following points:

${\bf (1)}$ \ \ \ explicit classification of all the weights $\mu$
participating in the same Weyl orbit ,

${\bf (2)}$ \ \ \ values of signatures $\epsilon(\mu)$ where
$\mu \in W(\Lambda^{++})$.

\noindent It will be seen that answers can be conveniently formulated in terms
of {\bf fundamental weights } {\bf [10]} \footnote{(*)}{ there must be no
confusion between fundamental weights $\mu_I$ and fundamental dominant
weights $\Lambda_\nu$'s or $\lambda_i$'s}
which we have previously introduced as being a convenient basis for weight
lattices of finite Lie algebras. We will see in this work that the use of
fundamental weights prove useful also for infinite dimensional Lie algebras.

\vskip3mm
\noindent{\bf {II. \ EXPLICIT CONSTRUCTION OF $A_N^{(1)}$ WEYL ORBITS }}
\vskip3mm

$A_N$ chain $(N=1,2, \dots)$ is in fact the essential key to understand the
structure of all other finite Lie algebras in the Cartan classification.
The same is also true for affine Lie algebras and hence we will proceed from
the now on in the framework of $A_N^{(1)}$ with the following Dynkin diagram:

$$ \hskip0.008cm  0  $$
$$  1 \hskip0.5cm 2 \hskip0.5cm 3 \ \ \ \ \dots \hskip0.5cm N-1 \hskip0.5cm N $$
We refer to excellent books of Kac {\bf[8]} and Humphreys {\bf[11]} for,
respectively, affine and finite Lie algebras though some notation which we
need in the sequel will also be given here. We define the fundamental
weights $\mu_I$ \ (I=1,2,.. N+1) of $A_N$ by
$$ \eqalign{
\mu_1 &\equiv {\bf \bar \lambda_1}  \cr
\mu_I &\equiv \mu_{I-1}-\alpha_{I-1} \ \ , \ \  I=2,3,..N+1 }
\eqno(II.1) $$
or, conversely, by
$$ {\bf \bar \lambda_i} = \sum_{j=1}^i \ \mu_j \ \ , \ \ i=1,2,.. N.
\eqno(II.2) $$
where ${\bf \bar \lambda_i}$'s are fundamental dominant weights and
$\alpha_i \ 's $ are simple roots of $A_N$. By definition, $A_N$ fundamental
weights are constrained by
$$ \sum_{I=1}^{N+1} \ \mu_I \equiv 0
\eqno(II.3) $$
and they provide us a symmetric scalar product
$$ (\mu_I,\mu_J) = \delta_{I,J} - {1 \over (N+1)} . \eqno(II.4) $$
The system of fundamental weights can be extended by adjoining the elements
$\Lambda_0$ , $\alpha_0$ and $\delta$ with the adopted symmetric scalar
products given below:
$$ \eqalign{
(\delta,\delta) = (\Lambda_0,\Lambda_0) &= 0  ,   \cr
(\Lambda_0,\delta) &= 1  ,                        \cr
(\mu_I,\delta) = 0  , (\mu_I,\Lambda_0) &= 0  .    }    \eqno(II.5) $$
It is thus seen that, in view of (II.4) and (II.5), the definitions
$$ \eqalign{
&\alpha_0 = \mu_{N+1} - \mu_1 + \delta  \cr
&\alpha_1 = \mu_1 - \mu_2               \cr
&\alpha_2 = \mu_2 - \mu_3               \cr
&\dots                                  \cr
&\alpha_N = \mu_N - \mu_{N+1}               } \eqno(II.6)  $$
give us a system of simple roots $\{ \alpha_{\nu} \ , \ \nu=0,1, \dots N \} $
for $A_N^{(1)}$ with the Dynkin diagram given above. Corresponding set
$\{ \Lambda_\nu \}$ of affine fundamental dominant weights can then be given by
$$ \Lambda_i = \Lambda_0 + {\bf \bar \lambda_i} \ \ ,
\ \ i=1,2, \dots N. \eqno(II.7) $$
together with $\Lambda_0$. In the specification of an affine weight $\Lambda$,
we have two central concepts of {\bf level} k and {\bf depth} M which are
defined by
$$ \eqalign{
k &\equiv (\lambda,\delta)   ,  \cr
M &\equiv (\lambda,\Lambda_0)  .  }  \eqno(II.8) $$
It is known that level can always be defined to be a positive integer and,
as a result of Schur lemma, it is constant for irreducible representations of
affine Lie algebras. Beyond affine Lie algebras, any value of \ k \ will
contribute in the same irreducible representation. The depth M, on the other
hand, can be conveniently defined to be zero for the principal dominant weight
$\Lambda^+$ of an irreducible affine representation $R(\Lambda^+)$ while it is
in general a non-positive integer for its sub-dominant weights.
\footnote{(*)}{ the concept of sub-dominant weights come from the
nomenclature of finite Lie algebras while they are called maximal
weights in the notation of Kac}

For any affine dominant weight $\Lambda^+$, its Weyl orbit $W(\Lambda^+)$ is
formed out of all weights $\sigma(\Lambda^+)$ where $\sigma$ is an affine
Weyl reflection for which the decomposition
$$ \sigma(\Lambda^+)= k \ \Lambda_0 + M \ \delta + {\bf \bar \mu}
\eqno(II.9) $$
is always valid on condition that
$$ \Lambda^+ \equiv k \ \Lambda_0 + \bar \Lambda^+
\eqno(II.10) $$
and also
$$ (\bar \mu,\bar \mu) - (\bar \Lambda^+,\bar \Lambda^+) = 2 \ k \ M
\eqno(II.11) $$
where both $\bar \mu$ and $\bar \Lambda^+$ are on the weight lattice of $A_N$.
We will show in a second work that a similar {\bf but not the same} relation
governs the Weyl orbits of infinite dimensional Lie algebras beyond affine ones.

As is also emphasized in the book of Kac, (II.11) is of central importance
for an explicit knowledge on the Weyl orbital structure of affine Lie algebras.
The most economical way to solve (II.11) is to recall in general that $A_N$
dominant weights $\bar \mu^+$ which respect (II.11) are to be specified by
$$ \bar \mu^+ = \bar \Lambda^+ + \sum_{i=1}^N r_i \ \alpha_i  .
\eqno(II.12) $$
where $r_i$'s are some non-negative integers. It is seen then that for each
and every order M of depth, there is only a finite number of sets
$\{ r_1,r_2, \dots r_N \}$. Let us define $\wp_M(\Lambda^+)$ as being the set
of $A_N$ dominant weights which correspond to these finite number of sets
$\{ r_1,r_2, \dots r_N \}$. This definition is of central importance due to
following lemma:

\hfill\eject

\vskip2mm
\noindent {\bf Lemma}
\vskip2mm
Any $A_N$ dominant weight ${ \bf \bar \mu^+}$ which solve (II.11) via (II.12)
is always contained in $\wp_M(\Lambda^+)$ if and only if $\Lambda^+$ is equal
to the one of affine fundamental dominant weights $\Lambda_0, \Lambda_1, \dots
\Lambda_N$. For any other
$$ \Lambda^+ \equiv \Lambda_{\nu_1} + \Lambda_{\nu_2} \ ,  $$
elements of $\wp_M(\Lambda^+)$ can be chosen uniquely from the set
$$ \wp_{M_1}( \Lambda_{\nu_1}) \oplus \wp_{M_2}(\Lambda_{\nu_2})  $$
on condition that (II.11) is also fulfilled. This can be repeated similarly
for any number of times.

\noindent We remark here that not all the weights of
$ \wp_{M_1}( \Lambda_{\nu_1}) \oplus \wp_{M_2}(\Lambda_{\nu_2})  $ fulfill
(II.11) and hence it is seen that this lemma completely solves the problem of
finding all the weights contributing in (I.1) or (I.4). One can now arrive
at the conclusions that

\vskip4mm

${\bf \bullet}$ \ \ \ the depth M can only be a non-positive integer,

${\bf \bullet \bullet}$ \ at each and every order M of depth, there can
only be a finite number of dominant weights ${\bf \bar \mu^+}$,

\noindent \ \ \ \ \ \ \ \ \ \ \ i.e. $ dim \wp_M(\Lambda^+) < \infty$ .

\vskip3mm

\noindent The following observation would however be of great help here.
Let us rewrite (II.11) for (II.12) together with
$$ {\bf \bar \Lambda^+} \equiv \sum_{i=1}^N \ p_i \ {\bf \bar \lambda_i}
\eqno(II.13) $$
where $p_i$'s are some non-negative integers. (II.11) will then turns out
to be the following algebraic equation:
$$ \sum_{i=1}^N r_i \ (r_i+p_i) - \sum_{i=1}^{N-1} r_i \ r_{i+1}=k \ M \ .
\eqno(II.14) $$

\noindent We finally conclude that:

\vskip2mm

\noindent \ \ \ Let $\Lambda$ be an affine dominant or strictly dominant
weight for which
$ \sigma(\Lambda)= k \ \Lambda_0 + M \ \delta + {\bf \bar \mu}  $
where $\sigma$ is an affine Weyl reflection. Let also
$ \sigma^+(\Lambda)= k \ \Lambda_0 + M \ \delta + {\bf \bar \mu^+}  $
where $\sigma^+$ is in general another affine Weyl reflection and
${\bf \bar \mu^+} \in \wp_M(\Lambda)$. Then the weights ${\bf \bar \mu}$
are in one-to-one correspondence with elements of finite Weyl orbits
$W({\bf \bar \mu^+}) . $ This in essence means that infinite summations over
affine Weyl groups of $A_N^{(1)}$ Lie algebras can be expressed in terms of
finite Weyl groups of $A_N$ Lie algebras. It is already known that,
{\bf in the fundamental weight basis}, actions of $A_N$ Weyl groups can
always be represented by ordinary permutations.

The second problem is to determine the signatures
$\epsilon(\mu)$ where $ \mu \in W(\Lambda^{++})$. In the light of following
specifications, this problem is to be solved in a quite non-trivial way. Let,
for $i=1,2, \dots N$, $n_i$'s be the set of non-negative integers and $s_i$'s
be the set of integers taking their values from the finite set
$\{0,1,2, \dots k \}$ for any positive integer k. As is given in above lemma,
let also that $\bar \mu^+$ be an $A_N$ dominant weight which solves
$$ (\bar \mu^+,\bar \mu^+) - (\bar \Lambda^{++},\bar \Lambda^{++}) =
2 \ k \ M  $$
in such a way that
$$ \sigma^{++}(\Lambda^{++})= k \ \Lambda_0 + M \ \delta + {\bf \bar \mu^+} $$
where $\sigma^{++}$ is an affine Weyl reflection. If one now considers the
decomposition
$$ {\bf \bar \mu^+} \equiv \sum_{i=1}^N (s_i + (N+1) \ n_i) \ \mu_i \ ,
\eqno(II.15) $$
the index defined on the right-hand side of the following expression will
give the correct signatures:
$$ \epsilon({\bf \bar \mu^+}) = \epsilon(s_1,s_2, \dots s_N) \
\prod_{i=1}^N \ (-1)^{n_i}     \eqno(II.16) $$
where $\epsilon(s_1,s_2, \dots s_N)$ \ 's are completely antisymmetric in
their indices and also
$$ \epsilon(s_1,s_2, \dots s_N) \equiv 1  \ \ , \ \
s_1 \geq s_2 \geq \dots \geq s_N \ \ . \eqno(II.17) $$
For any other Weyl conjugates ${\bf \bar \mu}$ of ${\bf \bar \mu^+}$ there
are additional contributions to (II.16) from the signatures of ordinary
permutations. This, in fact, shows how Schur functions enter in the
calculations of characters of $A_N$ Lie algebras.

In result, the problems mentioned in the introduction are now being solved.
In the next section, we will show that how can we use these results
in applications of Weyl-Kac character formula. As we will show in a subsequent
publication, it would be interesting to see how this also opens up a route
to systematic applications of Weyl-Kac character formula for infinite
dimensional Lie algebras beyond affine Lie algebras.

\vskip3mm
\noindent{\bf {III. \ CALCULATION OF STRING FUNCTIONS}}

In view of preceding section, we now explain how one makes use of the
existence of permutation weights in an explicit calculation of (I.1) and
(I.2). Most of our discussions, concerning $A_N$ Lie algebras, has been given
in detail in an unpublished work {\bf [12]} though some of them will also be
emphasized in the following.

For (I.1), we have two points to explain:

${\bf \bullet}$ \ \ \ the multiplicities $c_{\mu,\Lambda^+}(M)$ are invariant
under Weyl group actions, i.e. they have the same value for any $\mu$ having
the same depth M ,

${\bf \bullet \bullet}$ \ in the specialization $e^{\mu_I} = u_I$, characters
can be defined by
$$ChW({\bf \bar \mu^+}) = K_{q_1,q_2, \dots q_s}(u_1,u_2, \dots u_N)
\eqno(III.1) $$
for the whole Weyl orbit $W({\bf \bar \mu^+})$ of an $A_N$ dominant weight
$$ {\bf \bar \mu^+} \equiv \sum_{i=1}^s q_i \ \mu_i \ \ , \ \
q_1 \geq q_2 \geq \dots \geq q_s > 0 \eqno(III.2) $$
where $s \leq N$ and
$$ K_{q_1,q_2 \dots q_N}(u_1,u_2, \dots u_N) \equiv
\sum_{j_1,j_2, \dots j_s=1}^{N+1} \ (u_{j_1})^{q_1} \dots (u_{j_s})^{q_s} \ .
\eqno(III.3) $$
These polinomials can be reduced to class functions
$$ K_q(u_1,u_2 \dots u_N) \equiv \sum_{j=1}^{N+1} \ (u_j)^q \eqno(III.4) $$
via some reduction rules. Note here also that the parameters $u_I$ are
constrained by
$$ \prod_{I=1}^{N+1} u_I = 1 \eqno(III.5) $$
as a result of (II.3).

In ref.12, a detailed calculation of (I.2) is given for $A_N$ Lie algebras.
In view of decomposition (III.2) of $A_N$ dominant weights, we have, as a
result of some reduction procedure,
$$ A(\rho \ + \ {\bf \bar \mu^+}) = A(\rho) \
S_{q_1,q_2, \dots q_s}(x_1,x_2, \dots x_N) \eqno(III.6) $$
where
$$ A(\rho) = \prod_{j>i=1}^{N+1} \ (u_i-u_j) $$
is the Vandermonde determinant and
$ S_{q_1,q_2, \dots q_s}(x_1,x_2, \dots x_N)$'s
are Schur polinomials. The indeterminates $x_s$ are defined here by the
equivalence
$$ K_q(u_1,u_2 \dots u_N) \equiv q \ x_q \ \ , \ \
q=1,2 \dots N \ \ . \eqno(III.7) $$
(III.7) provides an inter-relation between the sets $ \{ u_1,u_2 \dots u_N \}$
and $ \{ x_1,x_2 \dots x_N \} $ of indeterminates and has crucial importance
to equate (I.1) and (I.2) in practical calculations. The reduction procedure
to express Schur polinomials in terms of {\bf classical Schur polinomials}
$S_q(x_1,x_2 \dots x_N)$ is known. We know that, for a fixed  value of N,
there are only N classical Schur polinomials
$S_q(x_1,x_2 \dots x_N)$ with $q=1,2 \dots N$ though one could also need
in general those of $q>N$. This arises the problem of extending classical
Schur polinomials $S_q(x_1,x_2 \dots x_N)$ for $q>N$ and these are the
so-called {\bf degenerated Schur polinomials}. To this end, it is shown in
ref.12 that the following reduction rules are sufficient:
$$ S_q = (-1)^{N+1} \ S_{q-N} - \sum_{r=1}^{N+1} \ S_r^*  \ S_{q-r} \ \ ,
\ \ q \geq N+1 \eqno(III.8) $$
where $S_q^*$ is obtained from $S_q$ under replacements $x_i \rightarrow -x_i$.

We will now show that the calculations for affine Lie algebras are to be
reduced to those of the finite ones if one uses permutation weights. To
proceed further, it will be more instructive to follow an explicit example.
A very typical one is, for instance, the level-2 representation
$R(\Lambda_0+\Lambda_1)$ of $A_5^{(1)}$. Its sub-dominants are given by
$$ \eqalign{
&M_0(\Lambda_0+\Lambda_1,\Lambda_0+\Lambda_1) = 0  ,   \cr
&M_0(\Lambda_2+\Lambda_5,\Lambda_0+\Lambda_1) = 1  ,   \cr
&M_0(\Lambda_3+\Lambda_4,\Lambda_0+\Lambda_1) = 2  .   }   \eqno(III.9) $$
We thus have, from (I.5), three string functions
$$ \eqalign{
&C_{\Lambda_0+\Lambda_1}(\Lambda_0+\Lambda_1) \equiv \sum_{M=0}^\infty q^M \
c_{\Lambda_0+\Lambda_1,\Lambda_0+\Lambda_1}(M) \ , \   \cr
&C_{\Lambda_0+\Lambda_1}(\Lambda_2+\Lambda_5) \equiv \sum_{M=1}^\infty q^M \
c_{\Lambda_2+\Lambda_5,\Lambda_0+\Lambda_1}(M) \ , \   \cr
&C_{\Lambda_0+\Lambda_1}(\Lambda_3+\Lambda_4) \equiv \sum_{M=2}^\infty q^M \
c_{\Lambda_3+\Lambda_4,\Lambda_0+\Lambda_1}(M) \ } \eqno(III.10)  $$
where $q \equiv e^{-\delta}$ is an indeterminate. Instead of multiplicities,
if one simply puts these string functions in (I.1) one obtains
$$ \eqalign{ Ch(\Lambda_0+\Lambda_1) =
&C_{\Lambda_0+\Lambda_1}(\Lambda_0+\Lambda_1) \ \sum_{M=0}^\infty q^M \
\sum_{{\bf \bar \mu^+} \in \wp(\Lambda_0+\Lambda_1,M)} \
K_{{\bf \bar \mu^+}}(u_1,u_2 \dots u_5) \  +    \cr
&C_{\Lambda_0+\Lambda_1}(\Lambda_2+\Lambda_5) \ \sum_{M=0}^\infty q^M \
\sum_{{\bf \bar \mu^+} \in \wp(\Lambda_2+\Lambda_5,M)} \
K_{{\bf \bar \mu^+}}(u_1,u_2 \dots u_5) \  +    \cr
&C_{\Lambda_0+\Lambda_1}(\Lambda_3+\Lambda_4) \ \sum_{M=0}^\infty q^M \
\sum_{{\bf \bar \mu^+} \in \wp(\Lambda_3+\Lambda_4,M)}  \
K_{{\bf \bar \mu^+}}(u_1,u_2 \dots u_5) \ .   }
\eqno(III.11)   $$
For any K=0,1,2, \dots , we formally define here
$$ \wp(\Lambda,K) \equiv \bigcup_{M=0}^K \wp_M(\Lambda) .  \eqno(III.12) $$
where $\bigcup$ means simple collection. We also find useful to adopt
the notation
$$ K_{q_1,q_2 \dots q_5}(u_1,u_2 \dots u_5) \equiv
K_{{\bf \bar \mu^+}}(u_1,u_2 \dots u_5)   \eqno(III.13)   $$
in the light of (III.2). As is emphasized above, the reduction rules given
in ref.12 allow us to reduce
$K_{{\bf \bar \mu^+}}(u_1,u_2 \dots u_5) $'s in terms of class functions
$ K_q(u_1,u_2 \dots u_5)$'s. In result, (III.11) can be expressed in the
following serie expansion:
$$ Ch(\Lambda_0+\Lambda_1) \simeq
\sum_{J=0}^K q^J \ LEFT_J(u_1,u_2 \dots u_5)  \ .   \eqno(III.14) $$

On the other hand, the simple factorization in (III.6) does not occur if one
wants to apply (I.2) for affine Lie algebras. Note that (III.6) is valid only
for finite Lie algebras and for infinite dimensional Lie algebras one could
only expect to obtain the string functions with only positive integer
coefficients, i.e. positive integer multiplicities. As in (III.13),
let us also assume that
$$ S_{q_1,q_2 \dots q_5}(x_1,x_2 \dots x_5) \equiv
S_{{\bf \bar \mu^+}}(x_1,x_2 \dots x_5)  \ .  \eqno(III.15)   $$
In view of (A.9) and (A.10), following results will then be obtained
from (I.4):
$$ \eqalign{
A(\tilde \rho,K) = &A(\rho) \ \sum_{{\bf \bar \mu^+} \in \wp(\tilde \rho,K)}
S_{{\bf \bar \mu^+}-\rho}(x_1,x_2 \dots x_5) \ ,  \cr
A(\tilde \rho+\Lambda_0+\Lambda_1,K) = &A(\rho) \
\sum_{{\bf \bar \mu^+} \in \wp(\tilde \rho+\Lambda_0+\Lambda_1,K)}
S_{{\bf \bar \mu^+}-\rho}(x_1,x_2 \dots x_5)       }  \eqno(III.16)  $$
where $A(\rho)$ is still the Vandermonde determinant defined above. It is
clear that this allows us to make the serie expansion
$$  { A(\tilde \rho+\Lambda_0+\Lambda_1,K) \over A(\tilde \rho,K ) } \simeq
\sum_{J=0}^K q^J \ RIGHT_J(x_1,x_2 \dots x_5)  \eqno(III.17) $$
which determine (I.2) up to any order K ( =0,1 \dots) .

The experienced reader knows here that actual calculations bring out severe
difficulties in practice. It will therefore be quite suitable to consider the
specialization
$$ u_1 = \kappa \ , \ u_2 = \kappa^{-1} \ , \ u_3 = u_4 = u_5 = 1
\eqno(III.18) $$
for which one has, as a result of (III.7),
$$ x_i = {4 \over i} + {\kappa \over i} + {\kappa^{-1} \over i}
\eqno(III.19) $$
for all values of i, i.e. $i \leq 5$ or $ i>5$. With this remark in mind, the
equality for the right-hand sides of (III.14) and (III.17) is always valid
for any fixed value of K and this is the Weyl-Kac formula which provides
the string functions in (III.10) up to order 9:
$$ \eqalign{
C_{\Lambda_0+\Lambda_1}(\Lambda_0+\Lambda_1) =&
1 + 10 \ q + 70 \ q^2 + 380 \ q^3 + 1740 \ q^4 + \cr
&7012 \ q^5 + 25585 \ q^6 + 86130 \ q^7 + 271225 \ q^8 + 807100 \ q^9 + \dots   \cr
C_{\Lambda_0+\Lambda_1}(\Lambda_2+\Lambda_5) =&q \ ( 2 + 22 \ q + 148 \ q^2 +
770 \ q^3 + 3382 \ q^4 + \cr
&13134 \ q^5 + 46382 \ q^6 + 151734 \ q^7 + 465894 \ q^8 + \dots )   \cr
C_{\Lambda_0+\Lambda_1}(\Lambda_3+\Lambda_4) =&q^2 \ (
5 + 50 \ q + 315 \ q^2 + 1550 \ q^3 + 6506 \ q^4 + 24320 \ q^5 +
83140 \ q^6 + 264460 \ q^7 + \dots )  \ .   } $$
All of these multiplicities will be obtained from equations which appear to be
coefficients of powers of the free parameter $\kappa$ in the Weyl-Kac formula
and it will be seen that the calculations need very restricted computer times,
say, in Mathematica {\bf [13]}.

\vskip3mm
\noindent{\bf {CONCLUSIONS}}
\vskip3mm

We have two conclusions one of which is for the present and the other is for a
future work. As is seen in above example, the equality of (III.14) and
(III.17) is valid for any order K of depth on condition that the corresponding
sets $\wp(\Lambda,K)$ of permutation weights are known for necessary
affine dominant weights $\Lambda$'s. And the whole machinery which developed
in this work for affine Lie algebras can be applied, with  only an appropriate
modification, beyond affine Lie algebras {\bf [14]}.

\hfill\eject

\vskip3mm
\noindent{\bf {REFERENCES}}
\vskip3mm

\leftline{[1] H.S.M.Coxeter, Discrete Groups Generated by Reflections,
Ann. Math., 35 (1934) 588  }
\leftline{ \ \ \ \ J.E.Humphreys, Reflection Groups and Coxeter Groups,
Cambridge Univ.Press (1990) }

\leftline{[2] J. Fuchs , Lectures on Conformal Field Theory and Kac-Moody Algebras,
hep-th/9702914   }

\leftline{[3] M.A.Olshanetsky and A.M.Perelemov, Lett.Math.Phys. 2 (1977) 7-12 }
\leftline{ \ \ \ \ \ \ \ \ \ \ \ \ \ \ \ \ \ \ \ \ \ \ \ \ \
\ \ \ \ \ \ \ \ \ \ \ \ \ \ \ \ \ \ \ \ \ \ \ \ \ \ \ \ \
Phys.Reports 94 (1983) 313 }

\leftline{[4] H.R.Karadayi and M.Gungormez, Summing over the Weyl Groups, math-ph/9812014 }

\leftline{[5] V.G.Kac and S.Peterson, Adv. Math., 53 (1984) 125-264}

\leftline{[6] H.Weyl, The Classical Groups, N.J. Princeton Univ. Press (1946)}

\leftline{[7] H.R.Karadayi and M.Gungormez, J.Phys.A:Math.Gen. 32 (1999) 1701-1707 }

\leftline{[8] V.G.Kac, Infinite Dimensional Lie Algebras, N.Y., Cambridge Univ. Press (1990)}
\leftline{ \ \ \ S.Kass, R.V.Moody, J.Patera and R.Slansky, Affine Lie Algebras,
Weight Multiplicities }
\leftline{ \ \ \ \ \ \ \ \ \ \ \ \ \ \ \ \ \ \ \ \ \ \ \ \ \ \ \ \ \ \ \ \ \ \
\ \ \ \ \ \ \ \ \ \ \ \ \ \ \ \ \ \ \ \ \ \ \
and Branching Rules, Univ. Calif.Press (1990)}

\leftline{[9] R.E.Borcherds, Generalized Kac-Moody Algebras, Jour.Algebra 115 (1988) 501-512 }

\leftline{[10] H.R.Karadayi, Anatomy of Grand Unifying Groups I and II , }
\leftline{ \ \ \ \ \ \ \ \ \ \ \ \ \ \ \ \ \ \ \ \ \ \ \ \ \
ICTP preprints(unpublished) IC/81/213 and 224}
\leftline{ \ \ \  H.R.Karadayi and M.Gungormez, Jour.Math.Phys., 38 (1997) 5991-6007}

\leftline{[11] J.E.Humphreys, Introduction to Lie Algebras and Representation Theory, N.Y., Springer-Verlag (1972)}

\leftline{[12] H.R.Karadayi, $A_N$ multiplicity rules and Schur functions, math-ph/9805009 (unpublished)}

\leftline{[13] S. Wolfram, Mathematica$^{TM}$, Addison-Wesley (1990) }

\leftline{[14] H.R.Karadayi and M.Gungormez, work in preparation }

\vskip3mm
\noindent{\bf {APPENDIX}}
\vskip3mm

In this appendix, the notation
$$ M \ \delta + \sum_{i=1}^N p_i \ \mu_i \equiv {(p_1,p_2 \dots p_N)}_{{-M}}
\eqno(A.1) $$
will be useful with the emphasis that all weights in any $\wp(\Lambda,K)$
will be of the same level with $\Lambda$. The complete data for permutation
weights which occur to find all the multiplicities encountered in the
calculation of $Ch(\Lambda_0+\Lambda_1)$ up to nineth order will be given.

All affiine fundamental dominant weights are of level-1 and their permutation
weigts are as in the following:
$$ \eqalign{ \wp(\Lambda_0,9) = \{ \ &(0,0,0,0,0)_0 \ , \      \cr
&(2,1,1,1,1)_1 \ , \                                           \cr
&(2,2,1,1,0)_2 \ , \                                           \cr
&(3,3,2,2,2)_3 \ , \ (2,2,2,0,0)_3 \ , \ (3,1,1,1,0)_3 \ , \   \cr
&(3,3,3,2,1)_4 \ , \ (4,2,2,2,2)_4 \ , \ (3,2,1,0,0)_4 \ , \   \cr
&(4,3,2,2,1)_5 \ , \                                           \cr
&(4,4,4,3,3)_6 \ , \ (3,3,3,3,0)_6 \ , \                       \cr
&(4,3,3,1,1)_6 \ , \ (3,3,0,0,0)_6 \ , \ (4,1,1,0,0)_6 \ , \   \cr
&(4,4,4,4,2)_7 \ , \ (5,4,3,3,3)_7 \ , \ (4,3,3,2,0)_7 \ , \   \cr
&(4,4,2,1,1)_7 \ , \ (5,2,2,2,1)_7 \ , \ (4,2,0,0,0)_7 \ , \   \cr
&(5,4,4,3,2)_8 \ , \ (4,4,2,2,0)_8 \ , \ (5,3,2,1,1)_8 \ , \   \cr
&(5,5,3,3,2)_9 \ , \ (4,4,3,1,0)_9 \ , \ (6,3,3,3,3)_9 \ , \
(5,3,2,2,0)_9 \ \}                 }           \eqno(A.2)      $$

$$ \eqalign{ \wp(\Lambda_1,9) = \{ \ &(1,0,0,0,0)_0 \ , \           \cr
&(2,2,1,1,1)_1 \ , \                                                \cr
&(2,2,2,1,0)_2 \ , \ (3,1,1,1,1)_2 \ , \                            \cr
&(3,3,3,2,2)_3 \ , \ (3,2,1,1,0)_3 \ , \                            \cr
&(3,3,3,3,1)_4 \ , \ (4,3,2,2,2)_4 \ , \ (3,2,2,0,0)_4 \ , \        \cr
&(4,3,3,2,1)_5 \ , \ (3,3,1,0,0)_5 \ , \ (4,1,1,1,0)_6 \ , \
(4,4,4,4,3)_6 \ , \                                                 \cr
&(4,4,2,2,1)_6 \ , \ (5,2,2,2,2)_6 \ , \ (4,2,1,0,0)_6 \ , \        \cr
&(5,4,4,3,3)_7 \ , \ (4,3,3,3,0)_7 \ , \ (4,4,3,1,1)_7 \ , \
(5,3,2,2,1)_7 \ , \                                                 \cr
&(5,4,4,4,2)_8 \ , \ (5,5,3,3,3)_8 \ , \                            \cr
&(4,4,3,2,0)_8 \ , \ (5,3,3,1,1)_8 \ , \ (4,3,0,0,0)_8 \ , \        \cr
&(5,5,4,3,2)_9 \ , \ (6,4,3,3,3)_9 \ , \                            \cr
&(5,3,3,2,0)_9 \ , \ (5,4,2,1,1)_9 \ , \ (5,1,1,0,0)_9 \ \} } \eqno(A.3) $$

$$ \eqalign{ \wp(\Lambda_2,9) = \{ \ &(1,1,0,0,0)_0 \ , \     \cr
&(2,2,2,1,1)_1 \ , \ (2,0,0,0,0)_1 \ , \                      \cr
&(2,2,2,2,0)_2 \ , \ (3,2,1,1,1)_2 \ , \                      \cr
&(3,3,3,3,2)_3 \ , \ (3,2,2,1,0)_3 \ , \                      \cr
&(4,3,3,2,2)_4 \ , \ (3,3,1,1,0)_4 \ , \ (4,1,1,1,1)_4 \ , \  \cr
&(4,3,3,3,1)_5 \ , \ (4,4,2,2,2)_5 \ , \ (3,3,2,0,0)_5 \ , \
(,4,2,1,1,0)_5 \ , \                                          \cr
&(4,4,4,4,4)_6 \ , \ (4,4,3,2,1)_6 \ , \ (5,3,2,2,2)_6 \ , \
(4,2,2,0,0)_6 \ , \                                           \cr
&(5,4,4,4,3)_7 \ , \ (5,3,3,2,1)_7 \ , \ (4,3,1,0,0)_7 \ , \  \cr
&(5,5,4,3,3)_8 \ , \ (4,4,3,3,0)_8 \ , \                      \cr
&(4,4,4,1,1)_8 \ , \ (5,4,2,2,1)_8 \ , \ (5,1,1,1,0)_8 \ , \  \cr
&(5,5,4,4,2)_9 \ , \ (4,4,4,2,0)_9 \ , \ (6,4,4,3,3)_9 \ , \  \cr
&(5,3,3,3,0)_9 \ , \ (5,4,3,1,1)_9 \ , \ (6,2,2,2,2)_9 \ , \
(5,2,1,0,0)_9  \  \}      }   \eqno(A.4)                     $$

$$ \eqalign{ \wp(\Lambda_3,9) = \{ \ &(1,1,1,0,0)_0 \ , \     \cr
&(2,2,2,2,1)_1 \ , \ (2,1,0,0,0)_1 \ , \                      \cr
&(3,2,2,1,1)_2 \ , \                                          \cr
&(3,3,3,3,3)_3 \ , \ (3,2,2,2,0)_3 \ , \ (3,3,1,1,1)_3 \ , \
(3,0,0,0,0)_3 \ , \                                           \cr
&(4,3,3,3,2)_4 \ , \ (3,3,2,1,0)_4 \ , \ (4,2,1,1,1)_4 \ , \  \cr
&(4,4,3,2,2)_5 \ , \ (4,2,2,1,0)_5 \ , \                      \cr
&(4,4,3,3,1)_6 \ , \ (3,3,3,0,0)_6 \ , \ (5,3,3,2,2)_6 \ , \
(4,3,1,1,0)_6 \ , \                                           \cr
&(5,4,4,4,4)_7 \ , \ (4,4,4,2,1)_7 \ , \ (5,3,3,3,1)_7 \ , \  \cr
&(5,4,2,2,2)_7 \ , \ (4,3,2,0,0)_7 \ , \ (5,1,1,1,1)_7 \ , \  \cr
&(5,5,4,4,3)_8 \ , \ (5,4,3,2,1)_8 \ , \ (5,2,1,1,0)_8 \ , \  \cr
&(5,5,5,3,3)_9 \ , \ (4,4,4,3,0)_9 \ , \ (6,4,4,4,3)_9 \ , \  \cr
&(4,4,1,0,0)_9 \ , \ (6,3,2,2,2)_9 \ , \ (5,2,2,0,0)_9 \  \}  }
\eqno(A.5)     $$
It is seen that $\wp(\Lambda_0,9)$ and $\wp(\Lambda_3,9)$ are real under
$A_N$ diagram automorphism while $\wp(\Lambda_4,9)$ ans $\wp(\Lambda_5,9)$
will be conjugates, respectively, to $\wp(\Lambda_2,9)$ and $\wp(\Lambda_1,9)$.

We know from above, we have three sub-dominant weight of level-2 for
$Ch(\Lambda_0+\Lambda_1)$. Among elements of the set
$ \wp_{M_1}(\Lambda_0) \oplus \wp_{M_2}(\Lambda_1) $,
the ones which fulfill (II.11) form
$$ \eqalign{ \wp(\Lambda_0+\Lambda_1,9) = \{ \ &(1,0,0,0,0)_0 \ , \  \cr
&(3,1,1,1,1)_1 \ , \                                                 \cr
&(4,3,2,2,2)_2 \ , \                                                 \cr
&(4,4,2,2,1)_3 \ , \ (5,2,2,2,2)_3 \ , \                             \cr
&(4,4,3,2,0)_4 \ , \ (5,5,3,3,3)_4 \ , \                             \cr
&(4,4,4,1,0)_5 \ , \ (5,4,2,2,0)_5 \ , \ (6,2,2,2,1)_5 \ , \         \cr
&(5,5,5,3,1)_6 \ , \ (6,3,2,2,0)_6 \ , \ (6,6,5,4,4)_6 \ , \
(7,3,3,3,3)_6 \ , \                                                  \cr
&(5,4,4,0,0)_7 \ , \ (6,4,2,1,0)_7 \ , \ (6,6,6,4,3)_7 \ , \
(7,6,4,4,4)_7 \ , \                                                  \cr
&(6,4,3,0,0)_8 \ , \ (6,6,6,5,2)_8 \ , \                             \cr
&(7,2,2,2,0)_8 \ , \ (7,5,3,3,1)_8 \ , \ (8,5,4,4,4)_8 \ , \         \cr
&(6,5,2,0,0)_9 \ , \ (7,6,6,4,2)_9 \ , \ (7,7,7,5,5)_9 \ , \
(8,6,4,4,3)_9  \ \} } \ .  \eqno(A.6)  $$

and similarly one has

$$ \eqalign{ \wp(\Lambda_2+\Lambda_5,9) = \{ \  &(2,2,1,1,1)_0 \ , \  \cr
&(3,2,1,1,0)_1 \ , \ (3,3,3,2,2)_1 \ , \                              \cr
&(3,3,1,0,0)_2 \ , \ (4,1,1,1,0)_2 \ , \ (4,3,3,2,1)_2 \ , \          \cr
&(4,3,3,3,0)_3 \ , \ (4,4,3,1,1)_3 \ , \ (5,3,2,2,1)_3 \ , \
(5,4,4,3,3)_3 \ , \                                                   \cr
&(5,1,1,0,0)_4 \ , \ (5,3,3,2,0)_4 \ , \                              \cr
&(5,4,2,1,1)_4 \ , \ (5,5,4,3,2)_4 \ , \ (6,4,3,3,3)_4 \ , \          \cr
&(5,4,3,1,0)_5 \ , \ (5,5,4,4,1)_5 \ , \ (5,5,5,2,2)_5 \ , \          \cr
&(6,3,2,1,1)_5 \ , \ (6,5,3,3,2)_5 \ , \ (6,5,5,5,4)_5 \ , \          \cr
&(5,5,2,1,0)_6 \ , \ (6,3,3,1,0)_6 \ , \ (6,4,1,1,1)_6 \ , \          \cr
&(6,5,4,3,1)_6 \ , \ (6,6,5,5,3)_6 \ , \ (7,4,3,3,2)_6 \ , \
(7,5,5,4,4)_6 \ , \                                                   \cr
&(5,5,3,0,0)_7 \ , \ (5,5,5,4,0)_7 \ , \ (6,5,5,2,1)_7 \ , \
(6,6,3,3,1)_7 \ , \                                                   \cr
&(7,2,2,1,1)_7 \ , \ (7,4,4,3,1)_7 \ , \ (7,5,3,2,2)_7 \ , \
(7,6,5,4,3)_7 \ , \                                                   \cr
&(6,5,1,1,0)_8 \ , \ (6,5,5,3,0)_8 \ , \ (7,3,2,1,0)_8 \ , \          \cr
&(7,5,4,2,1)_8 \ , \ (7,6,5,5,2)_8 \ , \ (7,6,6,3,3)_8 \ , \          \cr
&(7,7,4,4,3)_8 \ , \ (7,7,6,6,5)_8 \ , \ (8,3,3,3,2)_8 \ , \
(8,5,5,4,3)_8 \ , \                                                   \cr
&(6,6,5,1,1)_9 \ , \ (7,3,3,0,0)_9 \ , \ (7,4,1,1,0)_9 \ , \          \cr
&(7,5,4,3,0)_9 \ , \ (7,6,3,2,1)_9 \ , \ (7,7,5,4,2)_9 \ , \
(7,7,7,6,4)_9 \ , \                                                   \cr
&(8,4,3,3,1)_9 \ , \ (8,5,5,5,2)_9 \ , \ (8,6,5,3,3)_9 \ , \
(8,7,6,5,5)_9 \ \} }  \ . \eqno(A.7)     $$

$$ \eqalign{ \wp(\Lambda_3+\Lambda_4,9) = \{ \  &(2,2,2,1,0)_0 \ , \     \cr
&(3,2,2,0,0)_1 \ , \ (3,3,3,3,1)_1 \ , \                                 \cr
&(4,2,1,0,0)_2 \ , \ (4,4,4,4,3)_2 \ , \                                 \cr
&(4,3,0,0,0)_3 \ , \ (5,3,3,1,1)_3 \ , \ (5,4,4,4,2)_3 \ , \             \cr
&(5,2,0,0,0)_4 \ , \ (5,5,5,5,5)_4 \ , \ (6,4,4,3,2)_4 \ , \             \cr
&(5,5,1,1,1)_5 \ , \ (6,4,4,4,1)_5 \ , \ (6,5,4,2,2)_5 \ , \             \cr
&(6,1,0,0,0)_6 \ , \ (6,6,3,2,2)_6 \ , \ (7,4,4,2,2)_6 \ , \
(7,5,5,5,3)_6 \ , \                                                      \cr
&(6,5,4,4,0)_7 \ , \ (6,6,4,2,1)_7 \ , \ (7,3,1,1,1)_7 \ , \
(7,6,6,6,6)_7 \ , \                                                      \cr
&(6,6,4,3,0)_8 \ , \ (7,4,4,4,0)_8 \ , \ (7,6,2,2,2)_8 \ , \             \cr
&(7,7,5,3,3)_8 \ , \ (8,4,3,2,2)_8 \ , \ (8,6,6,6,5)_8 \ , \             \cr
&(6,6,5,2,0)_9 \ , \ (7,0,0,0,0)_9 \ , \                                 \cr
&(8,4,4,2,1)_9 \ , \ (8,5,2,2,2)_9 \ , \ (8,7,6,6,4)_9 \ \} \ . }
\eqno(A.8)  $$

The equivalent of $Ch(\Lambda_0+\Lambda_1)$ is provided by the Weyl-Kac formula.
For this, we first define affine Weyl vector $\tilde \rho$ as in the following:
$$\tilde \rho \equiv \sum_{\nu=0}^N \ \Lambda_\nu \ .  $$
By applying successively above lemma 6 times in any order, following
decomposition for $\wp(\tilde \rho,9)$ will then be obtained:
$$ \eqalign{ \ \wp(\tilde \rho,9) = \{ \ &(5,4,3,2,1)_0 \ ,         \cr
&(7,5,4,3,2)_1 \ ,                                                  \cr
&(8,5,4,3,1)_2 \ , \ (8,7,5,4,3)_2 \ ,                              \cr
&(9,5,4,2,1)_3 \ , \ (9,7,5,4,2)_3 \ , \ (9,8,7,5,4)_3 \ ,          \cr
&(10,5,3,2,1)_4 \ , \ (10,7,5,3,2)_4 \ ,                            \cr
&(10,8,7,5,3)_4 \ , \ (9,8,5,4,1)_4 \ , \ (10,9,8,7,5)_4 \ ,        \cr
&(11,4,3,2,1)_5 \ , \ (11,7,4,3,2)_5 \ , \ (11,8,7,4,3)_5 \ ,       \cr
&(10,8,5,3,1)_5 \ , \ (11,9,8,7,4)_5 \ , \ (10,9,7,5,2)_5 \ , \
(11,10,9,8,7)_5 \ ,                                                 \cr
&(10,9,5,2,1)_6 \ , \ (11,8,4,3,1)_6 \ ,                            \cr
&(11,9,7,4,2)_6 \ , \ (11,10,8,7,3)_6 \ , \ (10,9,8,5,1)_6 \ ,      \cr
&(13,11,10,9,8)_7 \ , \ (11,9,4,2,1)_7 \ , \ (11,10,7,3,2)_7 \ ,    \cr
&(13,10,9,8,5)_7 \ , \ (13,9,8,5,4)_7 \ , \ (13,8,5,4,3)_7 \ ,      \cr
&(13,5,4,3,2)_7 \ , \ (11,9,8,4,1)_7 \ , \ (11,10,9,7,2)_7 \ ,      \cr
&(14,11,10,9,7)_8 \ , \ (11,10,3,2,1)_8 \ , \ (11,10,8,3,1)_8 \ ,   \cr
&(14,10,9,7,5)_8 \ , \ (13,11,9,8,4)_8 \ , \ (14,9,7,5,4)_8 \ ,     \cr
&(13,10,8,5,3)_8 \ , \ (13,9,5,4,2)_8 \ , \ (14,7,5,4,3)_8 \ , \
(11,10,9,8,1)_8 \ ,                                                 \cr
&(15,11,10,8,7)_9 \ , \ (13,11,8,4,3)_9 \ , \ (13,10,5,3,2)_9 \ ,   \cr
&(14,13,11,10,9)_9 \ , \ (11,10,9,2,1)_9 \ , \ (15,10,8,7,5)_9 \ ,  \cr
&(14,11,9,7,4)_9 \ , \ (13,11,10,8,3)_9 \ , \ (14,10,7,5,3)_9 \ ,   \cr
&(15,8,7,5,4)_9 \ , \ (13,10,9,5,2)_9 \ , \ (14,5,4,3,1)_9  \}  \ .    }
(A,9) $$
One also has similarly
$$ \eqalign { \wp(\tilde \rho+\Lambda_0+\Lambda_1,9) =
\{ \ &(6,4,3,2,1)_0 \ ,\     \cr
&(0,0,0,0,0)_1 \ , \                                                      \cr
&(10,6,5,4,3)_2 \ , \                                                      \cr
&(11,6,5,4,2)_3 \ , \                                                      \cr
&(12,10,7,6,5)_4 \ , \ (12,6,5,3,2)_4 \ ,\                                 \cr
&(13,10,7,6,4)_5 \ , \ (13,11,9,7,6)_5 \ , \ (13,6,4,3,2)_5 \ , \          \cr
&(14,10,7,5,4)_6 \ , \ (14,11,9,7,5)_6 \ , \ (14,12,10,9,7)_6 \ , \
(14,5,4,3,2)_6 \ , \                                                       \cr
&(13,12,7,6,2)_7 \ , \ (15,10,6,5,4)_7 \ , \                               \cr
&(15,11,9,6,5)_7 \ , \ (15,12,10,9,6)_7 \ , \ (15,13,11,10,9)_7 \ , \      \cr
&(14,12,7,5,2)_8 \ , \ (14,13,9,7,3)_8 \ , \                               \cr
&(14,13,7,4,2)_9 \ , \ (15,12,6,5,2)_9 \ , \                               \cr
&(15,13,9,6,3)_9 \ , \ (15,14,10,9,4)_9 \ , \ (17,14,12,11,10)_9 \ , \  \cr
&(17,13,11,10,7)_9 \ , \ (17,12,10,7,6)_9 \ , \ (17,11,7,6,5)_9  \ \} \ .  }
\eqno(A.10)  $$

\end